\documentstyle[11pt,aaspp4,flushrt]{article}

\received{}
\accepted{}
\journalid{}{}
\articleid{}{}

\slugcomment{Accepted by the Astronomical Journal}

\lefthead{Crawford et al.}
\righthead{Polarization Properties of Radio Pulsars}

\newcommand\nudotdotdot{\ifmmode\stackrel{\bf \,...}{\textstyle
\nu}\else$\stackrel{\,...}{\textstyle \nu}$\fi}

\newcommand{\approxlt}{\mbox{$\;^{<}\hspace{-0.24cm}_{\sim}\;$}}
\newcommand{\approxgt}{\mbox{$\;^{>}\hspace{-0.24cm}_{\sim}\;$}}

\begin{document}

\title{Polarization Properties of Nine Southern Radio Pulsars}

\author{Fronefield Crawford\altaffilmark{1,2,3}, 
Richard N. Manchester\altaffilmark{4},
and Victoria M. Kaspi\altaffilmark{3,5,6}}

\altaffiltext{1}{Lockheed Martin Management and Data Systems, P.O. Box
8048, Philadelphia, PA 19101}

\altaffiltext{2}{Department of Physics, Haverford College, Haverford,
PA 19041}

\altaffiltext{3}{Department of Physics and Center for Space Research,
Massachusetts Institute of Technology, Cambridge, MA 02139}

\altaffiltext{4}{Australia Telescope National Facility, CSIRO,
P.O. Box 76, Epping, NSW 1710, Australia}

\altaffiltext{5}{Department of Physics, Rutherford Physics Building,
McGill University, 3600 University Street, Montreal, Quebec, H3A 2T8,
Canada}

\altaffiltext{6}{Alfred P. Sloan Research Fellow}

\begin{abstract}
We report on radio polarimetry observations for nine southern
pulsars. Six of the nine in the sample are young, with characteristic
ages under 100 kyr and high spin-down luminosities. All six show a
significant degree of linear polarization. We also confirm a
previously noticed trend in which the degree of linear polarization
increases with spin-down luminosity. Where possible we have used the
rotating-vector model of the pulsar emission geometry to fit the
observed position angle data.  Our fit for PSR J1513$-$5908
(B1509$-$58) in particular is useful for directly testing the
magnetospheric model of Melatos (1997) in combination with further
timing observations. For this pulsar we find that a magnetic
inclination angle greater than or equal to 60$^{\circ}$ is excluded at
the 3$\sigma$ level, and that the geometry suggested by the morphology
of an apparent bipolar X-ray outflow is marginally inconsistent with
the Melatos model. We also report on polarimetry of three older
pulsars: PSR J0045$-$7319, PSR J1627$-$4845, and PSR J1316$-$6232
(whose discovery we also report). Of these, only PSR J0045$-$7319
shows significant polarization.
\end{abstract}

\keywords{pulsars: general --- ISM: magnetic fields}

\section{Introduction \label{sec:intro}}

Polarization observations of radio pulsars are a useful way to probe
pulsar emission geometry, the radio emission process, and the
properties of the interstellar medium (ISM). The emission geometry of
a pulsar can be characterized by two angles: the angle between the
spin and magnetic dipole axes (the magnetic inclination angle,
$\alpha$), and the angle between the spin axis and the observer's line
of sight, $\zeta$. The difference between these two angles is the
impact parameter of the magnetic axis to the line of sight, $\beta$,
defined according to $\beta = \zeta - \alpha$. In the rotating-vector
model, the position angle (PA) of linearly polarized radiation follows
the projected direction of the magnetic field axis as the pulsar
rotates (\cite{rc69}). The PA, $\psi$, is expected to make a
characteristic S-shape as the axis swings across
the profile, and changes as a function of the pulse phase $\phi$
according to

\begin{equation} 
\tan (\psi - \psi_{0}) = \frac{\sin \alpha \sin (\phi -
\phi_{0})}{\sin \zeta \cos \alpha - \cos \zeta \sin \alpha \cos (\phi
- \phi_{0})} \, ,
\end{equation} 

\noindent
where $\psi_{0}$ is the projected direction of the rotation axis of
the pulsar and $\phi_{0}$ is the phase of maximum PA swing. By
rewriting the equation in terms of the directly measured $\beta$
rather than $\zeta$ and fitting the observed PA values to this model,
one can attempt to constrain these angles and understand the
orientation of the pulsar axes (e.g., \cite{lm88}).

In addition to elucidating pulsar magnetic and spin geometries,
polarization observations are useful for a variety of other
reasons. For example, Melatos (1997) has proposed a model of the
pulsar magnetosphere in which the pulsar and inner magnetosphere are
treated as a single perfectly conducting sphere rotating in a
vacuum. With three observable parameters (the period $P$, the period
derivative $\dot{P}$, and $\alpha$), the model predicts values for the
first and second braking indices, $n$ and $m$, which are defined as a
function of the pulsar frequency $\nu \equiv 1/P$ and its frequency
derivatives:

\begin{equation} 
n = \frac{\nu \ddot{\nu}}{\dot{\nu}^{2}} \, ,
\end{equation} 

\begin{equation}
m = \frac{\nu^{2} {\nudotdotdot}}{\dot{\nu}^{3}} \, .
\end{equation} 

\noindent
The model predicts values of $n$ for the Crab, PSR B0540$-$69, and PSR
J1513$-$5908 (B1509$-$58) which agree with the observed
values from timing data. The braking index measured for the recently
discovered PSR J1119$-$6127 is also in good agreement with the model
prediction (\cite{ckl+00}). To date, only the Crab pulsar
(\cite{lpg93}) and PSR J1513$-$5908 (\cite{kms+94}) have had
simultaneous measurements of $n$ and $m$ from a measured third period
derivative, and PSR J1513$-$5908 is the only pulsar for which this has
been done from absolute pulse numbering. Glitches and timing noise in
the Crab pulsar may never allow a direct estimate of its $m$,
though the Melatos (1997) prediction is consistent with current
estimates obtained indirectly by Lyne, Pritchard, \& Graham-Smith
(1993).  On the other hand, the accuracy of the spin parameters for
PSR J1513$-$5908 should increase over time. Therefore PSR J1513$-$5908
is the best pulsar currently known with which to test the Melatos
model. A previous estimate of $\alpha \sim 60^{\circ}$ was obtained
for PSR J1513$-$5908 by fitting the pulse profiles, and the relative
phase offsets of the peaks from radio, X-ray, and soft gamma-ray
observations, to an outer-gap emission model (\cite{ry95}).  Fitting
the observed polarization profile to the rotating vector model can in
principle provide a more direct and reliable constraint on $\alpha$.

There is a large body of literature in which the geometric
interpretation of pulsar polarization phenomenology and profile forms
is discussed. One interpretation (e.g., Rankin 1990\nocite{r90},
1993a\nocite{r93a}, 1993b\nocite{r93b}; \cite{gks93}; \cite{rr97};
\cite{wcl+99}) suggests that there is a clear split between a profile
component which arises from a central core beam and components which
stem from one or more hollow conal beams. Another interpretation,
largely due to Lyne \& Manchester (1988)\nocite{lm88}, suggests that
there is no fundamental difference in the origin of the different
profile forms and that all forms arise from ``patchy'' beams which
fill some fraction of an overall circular beam. In this scheme, a
gradual change in the emission characteristics with frequency and
pulse period, as well as viewing geometry, accounts for whether a
pulsar is core- or cone-dominated. In this paper we retain the core
and cone terminology and interpret our observed profiles in this
second context, though we do not pretend to claim that we
have resolved this contentious issue, which will require further
extensive work.

Polarimetric observations of young radio pulsars in general indicate
that they are typically more highly polarized than their older
counterparts (\cite{qml+95}; \cite{hkk98}). Polarization profiles for
some, especially young, pulsars have characteristics consistent with
their emission coming from part of a possibly wide conal beam
(\cite{lm88}; \cite{man97}). One consequence of this is that a
significant phase offset can be observed between the profile peak,
which occurs at one edge of the cone, and the point of maximum PA
swing, which traces the magnetic axis. If no significant phase offset
is seen, then a shallow PA swing would be present, implying a large
impact parameter. Conal beams are typically more highly linearly
polarized and have weak circular polarization, while core beams have
weak linear polarization and a sign reversal in the circular
polarization (\cite{lm88}; Rankin 1990\nocite{r90},
1993a\nocite{r93a}, 1993b\nocite{r93b}). It is not known why young
pulsars should exhibit strongly linearly polarized emission from conal
beams and why the degree of linear polarization should increase with
spin-down luminosity.

Polarization observations of pulsars also allow the determination of
rotation measures (RMs) which probe the Galactic magnetic field at a
variety of distances. The mean line-of-sight ISM magnetic field
strength can be estimated from measurements of the RM and dispersion
measure (DM) using

\begin{equation} 
\langle B_{||} \rangle = 1.232 \, \frac{\rm RM}{\rm DM} \, \mu{\rm G} \, ,
\end{equation} 

\noindent
where RM is measured in units of rad m$^{-2}$ and DM in units of pc
cm$^{-3}$ (\cite{mt77}). RM measurements can also help support
associations between young pulsars and supernova remnants
(SNRs). Traditionally this is done by measuring consistent ages and
distances for the pulsar and remnant. In most cases, however, there
are either significant discrepancies or large uncertainties in these
estimates. If the RM of the pulsar and the RM of the SNR at the
location of the pulsar are consistent with each other, then an
association would be supported.

We report on radio polarization observations for six young southern
radio pulsars. These pulsars all have characteristic ages $\tau_{c}
\equiv P/2\dot{P} <$ 100 kyr and spin-down luminosities $\dot{E}
\equiv 4\pi^{2} I \dot{P} / P^{3} > 10^{34}$ erg s$^{-1}$,
where $I$ is an assumed moment of inertia of 10$^{45}$ g cm$^{2}$.
Polarization results for most of the rest of the young pulsar
population can be found elsewhere (e.g., \cite{qml+95}; \cite{gl98};
\cite{mhq98}; \cite{hkk98}; \cite{wcl+99} and references therein). We
also report on polarimetry of three pulsars which are not particularly
young, but are interesting for other reasons: PSR J0045$-$7319 is in a
binary orbit with a B-star companion, PSR J1316$-$6232, whose
discovery we also report, is spatially coincident with a region of
extended radio emission, and PSR J1627$-$4845 is spatially coincident
with SNR G335.2+0.1. The properties for the pulsars in our sample are
listed in Table \ref{tbl-4}.

\section{Observations and Data Reduction}

All pulsars in our sample were observed with the Parkes 64-m radio
telescope in NSW, Australia, from 17 - 24 February 1997 at 1350 MHz,
except PSR J0045$-$7319 which was observed at 660 MHz. Follow-up
observations were conducted on several pulsars on 15 - 16 January 1998
at 660 and 2260 MHz. Table \ref{tbl-1} lists the parameters for the
observations. The details of the hardware setup and observing
technique are the same as those reported elsewhere (\cite{n94};
\cite{nms+97}; \cite{mhq98}). Observations were made in pairs at
orthogonal feed angles and summed to minimize the effects of
instrumental polarization. Residual instrumentation effects
contributed at most a few percent of the total intensity to the
measured Stokes parameters.

In cases where we did not have an initial RM estimate, an improper RM
correction to the data could cause the PAs from different frequency
channels to wind by more than one rotation and add
destructively. Therefore, we tried a large range of RMs, typically
$\pm 100$ DM rad m$^{-2}$, where DM is in units of pc cm$^{-3}$. In
most cases, one of the trial RMs yielded a significantly larger linear
polarization magnitude than the others. We used this RM as the
starting point for our RM convergence.

\section{Results and Discussion}

Seven of the nine pulsars (including all six of the young pulsars)
show linear polarization which is significant enough to permit an RM
estimate. The measured polarization parameters for the pulsars are
listed in Table \ref{tbl-2}, with $L$ and $V$ representing linearly
and circularly polarized intensity respectively. The polarization
profiles for the seven pulsars for which there was significant
polarization appear in Figure \ref{fig-1}. All of the profiles in this
figure are at 1350 MHz, except for PSR J0045$-$7319 and the additional
profile for PSR J1513$-$5908, both of which are at 660 MHz. Position
angles with more than 20$^{\circ}$ uncertainty were not plotted except
in the case of PSR J0045$-$7319 which was weak and therefore had an
imposed cutoff of 40$^{\circ}$ uncertainty. Table \ref{tbl-3} lists
the RMs and mean line-of-sight magnetic field strength estimates with
1$\sigma$ uncertainties for each. Total intensity pulse widths are
also included in Table \ref{tbl-3}.  In several cases the PA could be
measured over a significant fraction of the 1350 MHz profile with good
signal-to-noise. For these pulsars we attempted to fit the observed PA
swing over the profile to the rotating-vector model.

Below we outline the polarization results for each pulsar separately.

\subsection{PSR J0045$-$7319} 

PSR J0045$-$7319 was the first pulsar discovered in the Small
Magellanic Cloud (SMC) (\cite{mmh+91}) and is in a 51-day binary orbit
around a B1 class V star (\cite{kjb+94}). The wind from the B star,
however, is very tenuous and has been shown to contribute little to
the DM (\cite{ktm96}). This is consistent with our results which
indicate that during periastron passage (from 18 February 1997 to 23
February 1997) the RM remains small.

The 660 MHz polarization profile, compiled from a total of about 8
hours of observations made over 6 days, shows slight linear and more
significant circular polarization with large uncertainties in
each. The pulsar flux and polarization, however, are too weak to allow
us to say anything meaningful about the PA swing and emission
geometry. However, a secondary component to the right of the main peak
is likely one of the outlying profile components which is clearly
visible in higher frequency profiles. Since the majority of the
contribution to the DM ($\sim 75$\%) is likely to be from the SMC
itself, the measured RM (and mean line-of-sight magnetic field
strength) is strongly weighted toward the ISM of the SMC. The very low
value of $-$0.2 $\pm$ 0.3 $\mu$G for $\langle B_{||} \rangle$
indicates that this part of the SMC has magnetic fields which are
either tangled or largely perpendicular to the line of sight.

\subsection{PSR J1105$-$6107} 

PSR J1105$-$6107 is a young and energetic pulsar: its $\dot{E} = 2.5
\times 10^{36}$ erg s$^{-1}$, and
its characteristic age is 63 kyr (\cite{kbm+97}). As Figure
\ref{fig-1} shows, the 1350 MHz polarization profile shows two peaks
separated by $\sim 40^{\circ}$.  The profile is $\sim$ 100\% and
$\sim$ 80\% linearly polarized over the first and second peaks
respectively.  The second peak shows a slight increase in left-handed
circular polarization, though it remains weak. The high degree of
linear polarization and offset of the PA swing from the pulse peak
suggest that the emission is from the leading edge of a cone. The
occurrence of the maximum PA swing after the double peak indicates
that both components are likely part of the leading edge of the conal
beam (\cite{lm88}; \cite{man97}).  Our attempt to fit the PA data from
PSR J1105$-$6107 to the rotating-vector model did not yield a reliable
result and no formal constraints on the geometry of the pulsar can be
given.

\subsection{PSR J1316$-$6232} 

PSR J1316$-$6232 is a 343-ms pulsar with a large DM (983 pc cm$^{-3}$)
that was serendipitously discovered in the same survey targeting OB
stars in which PSR J1105$-$6107 was discovered (\cite{kbm+97}).
However, neither pulsar is associated with a target OB star; we report
the discovery of PSR J1316$-$6232 here.  Since its discovery, timing
observations have been carried out at the Parkes Observatory using a
hardware setup and analysis procedure which is described elsewhere
(\cite{dsb+98}).  A total of 90 observations spanning 733 days were
used at frequencies ranging from 1390 to 2000 MHz. Arrival times for
the observations were fitted using the TEMPO software
package\footnote{http://pulsar.princeton.edu/tempo}, with an rms
timing residual of $\sim$ 5 ms. The best-fit timing parameters for the
pulsar are given in Table \ref{tbl-6}.

The profile at 1350 MHz has a large asymmetric tail due to multipath
ISM scattering with a time constant $\tau_{s} \sim$ 150 ms.
Scattering may be smearing any linearly polarized emission, but at
2260 MHz, the scattering is significantly reduced and still no
significant linearly polarized emission is seen. The scattering at
2260 MHz is expected to be $\approxlt$ 20 ms, but the profile still
has a scattered appearance with $\tau_{s} \sim 50$ ms. It is likely
that there is an intrinsic component to the tail. The pulse width
remains large at high frequencies, with a duty cycle of $\sim$
20\%. No individual components can be resolved in the profile. A 4.85
GHz map of the region from the Parkes-MIT-NRAO Southern Survey
(\cite{cgw93}) shows a region of extended radio emission coincident
with the position of the pulsar. This may account for the large DM and
scattering if this region is in front of the pulsar. However, it is
unlikely that this would cause significant Faraday smearing since the
RM magnitudes that would be necessary (RM $> 10^{4}$ rad m$^{-2}$) far
exceed observed Galactic RM values (\cite{gld+99}). It is more likely
that PSR J1316$-$6232 is simply intrinsically unpolarized.

\subsection{PSR J1341$-$6220 (B1338$-$62)}

PSR J1341$-$6220 (B1338$-$62) is a young pulsar with $P = 193$ ms and
an age of 12 kyr (\cite{kmj+92}).  As shown in Figure \ref{fig-1}, at
1350 MHz the pulsar shows significant linear polarization (56\%) and
right-handed circular polarization (21\%). A previous estimate of the
linear polarization at 1400 MHz by Qiao et al. (1995) was somewhat
higher (80\%) but had a significantly larger uncertainty ($\sim$ 10\%)
than the results quoted here.  A scattering tail of length $\sim 8$ ms
is evident, which is not surprising considering the high DM (730 pc
cm$^{-3}$), and this has affected the polarization properties,
particularly in the trailing part of the pulse. Because of this we did
not attempt to fit the rotating-vector model to the PA data.

\subsection{PSR J1513$-$5908 (B1509$-$58)}

PSR J1513$-$5908 (B1509$-$58) was first discovered as an X-ray source
(\cite{sh82}) and was subsequently found to be a radio pulsar
(\cite{mtd82}). Timing observations showed that it has a large
$\dot{E} = 2 \times 10^{37}$ erg s$^{-1}$ and a small characteristic
age ($\tau_{c} \sim 1.5$ kyr). The pulsar is located near SNR
G320.4$-$1.2; the two are almost certainly associated
({\cite{gbm+99}).

Profiles for PSR J1513$-$5908 at both 660 and 1350 MHz are shown in
Figure \ref{fig-1}. At 660 MHz the profile is broad and probably
affected by interstellar scattering. There is good evidence at 1350
MHz for a weak component preceding the main pulse by $\sim$
50$^{\circ}$ of longitude. At both frequencies the emission is almost
completely linearly polarized, with significant right-handed circular
polarization as well. The high degree of linear polarization, shallow
PA swing, and absence of a sign change in the circular polarization
over the profile all support the notion that the emission is from the
grazing edge of a conal beam component.

There is no significant falloff in the degree of linear polarization
between 660 and 1350 MHz; if a falloff occurs, it is at higher
frequencies (\cite{mtt+73}; \cite{mt77}). The unchanging linear
polarization supports the observation of von Hoensbroech et al.
(1998) that high-$\dot{E}$ pulsars do not suffer as significantly from
depolarization effects at high frequencies as do low-$\dot{E}$
pulsars.

The RM estimates from the data at the two frequencies ($+211 \pm 5$
rad m$^{-2}$ for 660 MHz and $+215 \pm 2$ rad m$^{-2}$ for 1350 MHz)
are consistent with each other and are consistent with the RM of SNR
G320.4$-$1.2 of $+210 \pm 30$ rad m$^{-2}$ at the location of the
pulsar (\cite{gbm+99}). This supports the association between the
pulsar and the remnant, considering that the RM in different parts of
the SNR can vary by hundreds of rad m$^{-2}$.

As mentioned in \S 1, the magnetospheric model of Melatos
(1997) can be tested with an accurate determination of the emission
geometry of PSR J1513$-$5908. In particular, an estimate of $\alpha$
can be used along with $P$ and $\dot{P}$ to make predictions for the
braking indices $n$ and $m$. These predictions can be compared to the
measured values of $n$ and $m$ from pulsar timing (\cite{kms+94}).
From our fit of the 1350 MHz data to the rotating-vector model, values
of $\alpha \geq 60^{\circ}$ are excluded at the 3$\sigma$ level. An
estimate of $\sim 60^{\circ}$ was suggested from a fit of the
high-energy pulse profile to an outer-gap model (\cite{ry95}). Our
result suggests that this value is unlikely.

For an inclination angle range of $\alpha < 60^{\circ}$, the Melatos
(1997) model predicts that the first braking index should be $n <
2.92$ for PSR J1513$-$5908. The observed value of 2.837 $\pm$ 0.001
(\cite{kms+94}) falls easily in this range. For $\alpha < 60^{\circ}$,
the second braking index is predicted by the model to be $m <
14.0$. Currently the best measured value for this parameter is $m =
14.5 \pm 3.6$ (\cite{kms+94}), which partially overlaps the predicted
range. Therefore the Melatos model cannot be directly tested using
current data, but this should be possible in the future.

The swing of the observed PA data is shallow, implying a large impact
parameter. However, fitted values of $\alpha$ and $\zeta$ are highly
covariant because of the limited longitude extent of the observed
pulse profile; for $\alpha = 60^{\circ}$, $\zeta = 130^{\circ}$ and
for $\alpha = 45^{\circ}$, $\zeta = 100^{\circ}$. Therefore the
observations are consistent with the estimate of $\zeta \approxgt
70^{\circ}$ from Brazier \& Becker (1997)\nocite{bb97}, obtained by
interpreting the X-ray morphology of the surrounding region as the
result of a bipolar outflow from the pulsar with an equatorial
torus. However, by imposing the restriction that $\zeta > 70^{\circ}$
in our fit, we find that $\alpha > 30^{\circ}$ at the 3$\sigma$ level,
which corresponds to a prediction of $n > 2.86$ in the Melatos
model. This is marginally inconsistent with the observed value of $n =
2.837 \pm 0.001$ (\cite{kms+94}).

\subsection{PSR J1627$-$4845} 

PSR J1627$-$4845 was one of two new pulsars discovered in a targeted
search of southern supernova remnants (\cite{kmj+96}) and is spatially
coincident with SNR G335.2+0.1. However, the discrepancy in the ages
of the objects argues that the spatial association is only a chance
superposition. We were unable to detect significant linear
polarization or determine a rotation measure for this pulsar at 1350
MHz.  It is likely that the intrinsic polarization is low as might be
expected for an old pulsar.

\subsection{PSR J1646$-$4346 (B1643$-$43)}  

PSR J1646$-$4346 (B1643$-$43) is a 232-ms pulsar with a characteristic
age of 32 kyr
(\cite{jml+95}). As Figure \ref{fig-1} shows, our polarization profile
is noisy but has significant linear polarization (45\%) and negligible
circular polarization. There is also a slight profile asymmetry
probably from multipath scattering ($\tau_{s} \sim 11$ ms).  The
estimated RM of $-65 \pm 17$ rad m$^{-2}$ implies $\langle B_{||}
\rangle = -0.16 \pm 0.04$ $\mu$G, which is consistent with Galactic
magnetic field strengths.

\subsection{PSR J1730$-$3350 (B1727$-$33)} 

PSR J1730$-$3350 (B1727$-$33) has a period of 139 ms and is young,
with a characteristic age of 26 kyr (\cite{jml+95}). The polarization
profile at 1350 MHz shown in Figure \ref{fig-1} shows significant
linear polarization (56\%) with weak circular polarization. This is
consistent with the results of Gould \& Lyne (1998) who find the
degree of linear polarization to be 48\% at 1400 MHz. There is also a
significant scattering tail ($\tau_{s} \sim 7$ ms) from multipath
scattering, and the linear polarization in the trailing part of the
profile is affected by this. The rapid PA swing in the leading part of
the profile indicates a small impact parameter. The intrinsic profile
probably has a somewhat higher polarization with a rapid PA swing over a
greater range than that observed. A fit to the PA data preceding the
profile peak (where scattering has little effect) indicates that $|
\beta | < 5^{\circ}$ at the 3$\sigma$ confidence level.  The measured
RM of $-$142 $\pm$ 5 rad m$^{-2}$ implies $\langle B_{||} \rangle =
-0.68 \pm 0.02$ $\mu$G. This RM is consistent with RM measurements of
other pulsars in the region with comparable DMs.

\subsection{PSR J1801$-$2306 (B1758$-$23)} 

PSR J1801$-$2306 (B1758$-$23) was discovered in a search for pulsars
directed at Galactic objects (\cite{mdt85}). This pulsar was found to
be near SNR W28, which led to speculation that the pulsar and remnant
were associated (\cite{mkj+91}). The pulsar's characteristic age of 58
kyr is consistent with the estimated age of W28 of between 35 and 150
kyr (\cite{klm+93}), but there is a discrepancy between the remnant
distance of $\sim$ 3 kpc (\cite{fkv93}) and the pulsar distance, which
is estimated from its DM to be $\sim$ 13 kpc (\cite{tc93}). Attempts
have been made to resolve this discrepancy by introducing an H{\sc ii}
region in the line of sight which could contribute to the bulk of the
dispersion (\cite{fkv93}). However, there is no clear evidence for
such an H{\sc ii} region (\cite{klm+93}).

The pulse profile at 1350 MHz shows a smaller degree of linear
polarization (14\%) than the other young pulsars we have observed, and
is consistent with the estimate of Gould \& Lyne (1998), who find that
the linear polarization is 15\% at both 1400 and 1600 MHz. The
relatively small degree of polarization could be due to depolarization
from profile smearing from the strong scattering ($\tau_{s} \sim
100$ ms) and from differential Faraday rotation along the different
scattering paths in a region of high magnetic field and electron
density (such as an H{\sc ii} region or SNR). At 2260 MHz we find that the
linear polarization is low, less than a few percent, despite the
significant decrease in the scattering ($\tau_{s} \sim 15$ ms). The RM
can be estimated from the 1350 MHz data and is large and negative
(Table \ref{tbl-3}). We conclude that PSR J1801$-$2306 is intrinsically
less polarized than the other young pulsars in our sample. This is
consistent with the trend noted in Figure \ref{fig-2} since PSR
J1801$-$2306 has a smaller $\dot{E}$ than these pulsars (see below).

\subsection{Linear Polarization vs. Spin-down Luminosity}

By comparing the measured degree of linear polarization of our young
pulsars with their spin-down luminosities, $\dot{E}$, we confirm a trend
previously noticed by several authors (\cite{wml+93}; \cite{qml+95};
\cite{hkk98}) in which linear polarization increases with spin-down
luminosity. Figure \ref{fig-2} shows measured linear polarization as a
function of spin-down luminosity for our six young pulsars at 1350 MHz
(open circles) and for 278 pulsars from Gould and Lyne (1998) at 1400
MHz (dots). Our young pulsars are concentrated at the high-$\dot{E}$
end of the plot while the Gould and Lyne pulsars span an $\dot{E}$
range from $10^{30}$ to $10^{37}$ erg s$^{-1}$.  The figure shows a
trend in which pulsars with $\dot{E} > 10^{34}$ erg s$^{-1}$ tend to
have stronger linear polarization as $\dot{E}$ increases. The six
young pulsars in our sample show this trend clearly.

von Hoensbroech et al. (1998) noticed a stronger correlation in a
sample of 32 pulsars at a much higher frequency (4.9 GHz). Their trend
extends down to about $10^{32}$ erg s$^{-1}$.  Gould \& Lyne (1998)
find no such correlation at 400 MHz for any $\dot{E}$ range, and von
Hoensbroech et al. conclude that depolarization effects at high
frequencies affect low-$\dot{E}$ pulsars much more strongly than they
do high-$\dot{E}$ ones. Our 1350 MHz results confirm these trends.

\section{Conclusions} 

We have reported on radio polarimetry observations of nine southern
radio pulsars. Significant polarization was detected in seven of the
nine, including all six of the young pulsars in the sample. The two
pulsars with no detectable polarization (PSRs J1316$-$6232 and
J1627$-$4845) are older, with characteristic ages $\tau_{c} > 1$ Myr.
With the exception of PSR J1801$-$2306, all of our young pulsars
exhibit a high degree of linear polarization and a low degree of
circular polarization. This is consistent with the suggestion of
Manchester (1996) that young pulsars exhibit partial conal
emission. The rotating-vector model fit of the PSR J1513$-$5908 data
indicates that a magnetic inclination angle of $\alpha \geq
60^{\circ}$ is excluded at the 3$\sigma$ confidence level. This result
suggests that the value of $\alpha \sim 60^{\circ}$ estimated for this
pulsar by Romani \& Yadigaroglu (1995) is possible but
unlikely. Imposing the restriction in the fit that $\zeta >
70^{\circ}$ from Brazier \& Becker (1997) implies that $\alpha >
30^{\circ}$ at the 3$\sigma$ level. For $\alpha > 30^{\circ}$, the
predicted braking index $n > 2.86$ from the Melatos (1997) model is
marginally inconsistent with the observed value of $n = 2.837 \pm
0.001$ (\cite{kms+94}). Thus the predictions of the Melatos model and
the limit of $\zeta \approxgt 70^{\circ}$ derived by Brazier \& Becker
(1997) are marginally inconsistent with each other.  Finally, we confirm at
1350 MHz a positive correlation between degree of linear polarization
and spin-down luminosity.

\acknowledgments

We thank N. D'Amico for his assistance in the search that found PSR
J1316$-$6232, and M. Bailes and R. Pace for help with the subsequent
timing. FC thanks D. Fox for helpful discussions regarding model
fitting. The Parkes radio telescope is part of the Australia
Telescope, which is funded by the Commonwealth of Australia for
operation as a National Facility operated by CSIRO. VMK was supported
in part by an Alfred P. Sloan Research Fellowship, NSF Career Award
(AST-9875897), and NSERC grant RGPIN 228738-00.

%\begin{table}
%\dummytable\label{tbl-4}
%\end{table}

%\begin{table}
%\dummytable\label{tbl-1}
%\end{table}

%\begin{table}
%\dummytable\label{tbl-2}
%\end{table}

%\begin{table}
%\dummytable\label{tbl-3}
%\end{table}

%\begin{table}
%\dummytable\label{tbl-6}
%\end{table}

\begin{deluxetable}{cccccccccc}
\footnotesize
\tablecaption{Pulsar Characteristics \label{tbl-4}}
%\tablewidth{0pt}
\tablehead{
\colhead{PSR} &
\colhead{P} &
\colhead{DM} &
\colhead{$\log \tau_{c}$ \tablenotemark{a}} &
\colhead{$\log B$ \tablenotemark{b}} &
\colhead{$\log \dot{E}$ \tablenotemark{c}}  \\ 
\colhead{} & 
\colhead{(sec)} &
\colhead{(pc cm$^{-3}$)} &
\colhead{(yr)} & 
\colhead{(G)} & 
\colhead{(erg s$^{-1}$)}
}
\startdata
J0045$-$7319 & 0.926 &  105 & 6.5 & 12.3  & 32.3 \nl
J1105$-$6107 & 0.063 &  271 & 4.8 & 12.0  & 36.4 \nl
J1316$-$6232 & 0.343 &  983 & 6.0 & 12.2  & 33.7 \nl
J1341$-$6220 & 0.193 &  730 & 4.1 & 12.9  & 36.2 \nl
J1513$-$5908 & 0.151 &  255 & 3.2 & 13.2  & 37.3 \nl
J1627$-$4845 & 0.612 &  557 & 6.4 & 12.2  & 32.8 \nl
J1646$-$4346 & 0.232 &  490 & 4.5 & 12.7  & 35.6 \nl
J1730$-$3350 & 0.139 &  257 & 4.4 & 12.5  & 36.1 \nl
J1801$-$2306 & 0.415 & 1074 & 4.8 & 12.8  & 34.8 \nl
\enddata
 
\tablenotetext{a}{Characteristic age $\tau_{c} \equiv P/2\dot{P}$.} 
\tablenotetext{b}{Surface dipole magnetic field strength $B \equiv 3.2 \times 10^{19} (P\dot{P})^{1/2}$ G.}
\tablenotetext{c}{Spin-down luminosity $\dot{E} \equiv 4\pi^{2}I\dot{P
}/P^{3}$.}
\end{deluxetable}

\begin{deluxetable}{ccccccc}
\footnotesize
\tablecaption{Observing Parameters \label{tbl-1}}
%\tablewidth{0pt}
\tablehead{
\colhead{PSR} & 
\colhead{Date} &
\colhead{Freq} &
\colhead{T$_{int}$ \tablenotemark{a}} & 
\colhead{N$_{bins}$ \tablenotemark{b}} &
\colhead{BW \tablenotemark{c}} &  
\colhead{N$_{chan}$ \tablenotemark{d}} \\
\colhead{} & 
\colhead{} &
\colhead{(MHz)} &
\colhead{(min)} & 
\colhead{} &
\colhead{(MHz)} &  
\colhead{} 
}
\startdata
J0045$-$7319 & Feb 1997 & 660  & 464 & 256 & 32  & 8  \nl 
J1105$-$6107 & Feb 1997 & 1350 & 66  & 512 & 128 & 16 \nl 
J1316$-$6232 & Feb 1997 & 1350 & 72  & 128 & 128 & 64 \nl 
             & Jan 1998 & 2264 & 177 & 256 & 128 & 8  \nl
J1341$-$6220 & Feb 1997 & 1350 & 24  & 512 & 128 & 32 \nl 
J1513$-$5908 & Jan 1998 & 660  & 372 & 128 & 32  & 8  \nl 
             & Feb 1997 & 1350 & 192 & 256 & 128 & 32 \nl
J1627$-$4845 & Feb 1997 & 1350 & 24  & 512 & 128 & 32 \nl 
J1646$-$4346 & Feb 1997 & 1350 & 27  & 512 & 128 & 32 \nl 
J1730$-$3350 & Feb 1997 & 1350 & 30  & 512 & 128 & 16 \nl 
J1801$-$2306 & Feb 1997 & 1350 & 129 & 128 & 128 & 64 \nl 
             & Jan 1998 & 2264 & 372 & 256 & 128 & 8  \nl
\enddata

\tablenotetext{a}{Total integration time.}
\tablenotetext{b}{Number of bins in profile.}
\tablenotetext{c}{Observing bandwidth.}
\tablenotetext{d}{Number of frequency channels.}  
 
\end{deluxetable}

\begin{deluxetable}{ccccccc}
\footnotesize
\tablecaption{Pulsar Polarization Parameters \label{tbl-2}}
%\tablewidth{0pt}
\tablehead{
\colhead{PSR} & 
\colhead{Freq} &
\colhead{S \tablenotemark{a}} &
\colhead{$\langle$L$\rangle$/S \tablenotemark{b}} &
\colhead{$\langle$V$\rangle$/S \tablenotemark{c}} &
\colhead{$\langle |$V$| \rangle$/S \tablenotemark{d}} &
\colhead{error \tablenotemark{e}} \\
\colhead{} & 
\colhead{(MHz)} &
\colhead{(mJy)} &
\colhead{(\%)} &
\colhead{(\%)} &
\colhead{(\%)} &
\colhead{(\%)} 
}
\startdata
J0045$-$7319 & 660  & 0.7   & 11   & $-$17 & 27   & 7    \nl  
J1105$-$6107 & 1350 & 1.0   & 89   & $+$9  & 12   & 2    \nl 
J1316$-$6232 & 1350 & 1.8   & $-$  & $-$   & $-$  & $-$  \nl 
             & 2264 & 0.6   & $-$  & $-$   & $-$  & $-$  \nl
J1341$-$6220 & 1350 & 2.3   & 56   & $-$21 & 22   & 2    \nl 
J1513$-$5908 & 660  & 2.6   & 97   & $-$23 & 25   & 7    \nl    
             & 1350 & 1.3   & 94   & $-$18 & 18   & 2    \nl
J1627$-$4845 & 1350 & 0.5   & $-$  & $-$   & $-$  & $-$  \nl 
J1646$-$4346 & 1350 & 0.8   & 45   & $+$6  & 13   & 5    \nl 
J1730$-$3350 & 1350 & 3.3   & 56   & $-$5  & 7    & 1    \nl 
J1801$-$2306 & 1350 & 6.2   & 14   & $+$12 & 13   & 2    \nl
             & 2264 & 2.1   & $-$2 & $+$7  & 12   & 4    \nl
\enddata

\tablenotetext{a}{Mean flux density.} 
\tablenotetext{b}{Fractional linear polarization of on-pulse bins.} 
\tablenotetext{c}{Fractional circular polarization of on-pulse bins. Positive values are left-circularly polarized.}
\tablenotetext{d}{Fractional value of absolute circular polarization of on-pulse bins.}
\tablenotetext{e}{$2 \sigma$ uncertainty in polarization values. The uncertainty
due to instrumental effects is at most a few percent.}
 
\end{deluxetable}

\begin{deluxetable}{cccccc}
\footnotesize
\tablecaption{Measured Pulse Widths and Rotation Measures \label{tbl-3}}
%\tablewidth{0pt}
\tablehead{
\colhead{PSR} & 
\colhead{Freq} &
\colhead{W$_{50}$ \tablenotemark{a}} &
\colhead{W$_{10}$ \tablenotemark{b}} & 
\colhead{RM} &
\colhead{$\langle$B$_{||}\rangle$ \tablenotemark{c}} \\
\colhead{} & 
\colhead{(MHz)} &
\colhead{(ms)} &
\colhead{(ms)} & 
\colhead{(rad m$^{-2}$)} &
\colhead{($\mu$G)} 
}
\startdata
J0045$-$7319 & 660  &  20 & 220 & $-$14   $\pm$ 27  & $-$0.2 $\pm$ 0.3     \nl 
J1105$-$6107 & 1350 &   4 &  10 & $+$166  $\pm$ 3   & $+$0.76 $\pm$ 0.01   \nl  
J1316$-$6232 & 1350 & 119 & 180 &     $-$           &         $-$          \nl 
             & 2264 &  81 & 115 &     $-$           &         $-$          \nl
J1341$-$6220 & 1350 &  10 &  45 & $-$946  $\pm$ 7   & $-$1.60 $\pm$ 0.01   \nl  
J1513$-$5908 & 660  &  36 &  75 & $+$211  $\pm$ 5   & $+$1.02 $\pm$ 0.03   \nl  
             & 1350 &  14 &  40 & $+$215  $\pm$ 2   & $+$1.046 $\pm$ 0.008 \nl
J1627$-$4845 & 1350 &  32 & 170 &     $-$           &         $-$          \nl 
J1646$-$4346 & 1350 &  12 &  70 & $-$65   $\pm$ 17  & $-$0.16 $\pm$ 0.04   \nl 
J1730$-$3350 & 1350 &   7 &  20 & $-$142  $\pm$ 5   & $-$0.68 $\pm$ 0.02   \nl 
J1801$-$2306 & 1350 &  88 & 265 & $-$1156 $\pm$ 19  & $-$1.33 $\pm$ 0.02   \nl
             & 2264 &  19 &  80 & $-$               & $-$                  \nl
\enddata
 
\tablenotetext{a}{Width at which the measured pulse reaches 50\% of 
its peak (FWHM), with uncertainty 1 ms.}
\tablenotetext{b}{Width at which the measured pulse reaches 10\% of its peak,
with uncertainty 5 ms.}
\tablenotetext{c}{Negative values correspond to the magnetic field pointing 
away from observer.}
 
\end{deluxetable}

\begin{deluxetable}{ll}
\footnotesize
\tablecaption{Astrometric and Spin Parameters for PSR J1316$-$6232 \label{tbl-6}}
%\tablewidth{0pt}
\tablehead{
\colhead{Parameter} & 
\colhead{Value \tablenotemark{a}} 
} 
\startdata
Right Ascension, $\alpha$ (J2000) & $13^{\rm h} 16^{\rm m} 46^{\rm s}.3$(2) \nl 
Declination, $\delta$ (J2000)     & $-62^{\circ} 32^{\prime} 12^{\prime \prime}.2$(5) \nl
Galactic Latitude, $l$            & 305$^{\circ}$.85 \nl
Galactic Longitude, $b$           & +0$^{\circ}$.19 \nl
Period, $P$                       &  0.34282539844(6) s \nl
Period Derivative, $\dot{P}$      &  5.297(2) $\times 10^{-15}$ \nl
Dispersion Measure, DM            &  983(2) pc cm$^{-3}$ \nl
Epoch of Period                   &  MJD 49800.0000 \nl
Timing Span                       &  MJD 49729 to MJD 50462 \nl
Timing Residual                   &  4.67 ms                \nl
\enddata

\tablenotetext{a}{Uncertainties in parentheses apply to the last digit.}

\end{deluxetable}

\newpage

\begin{figure}
\plotone{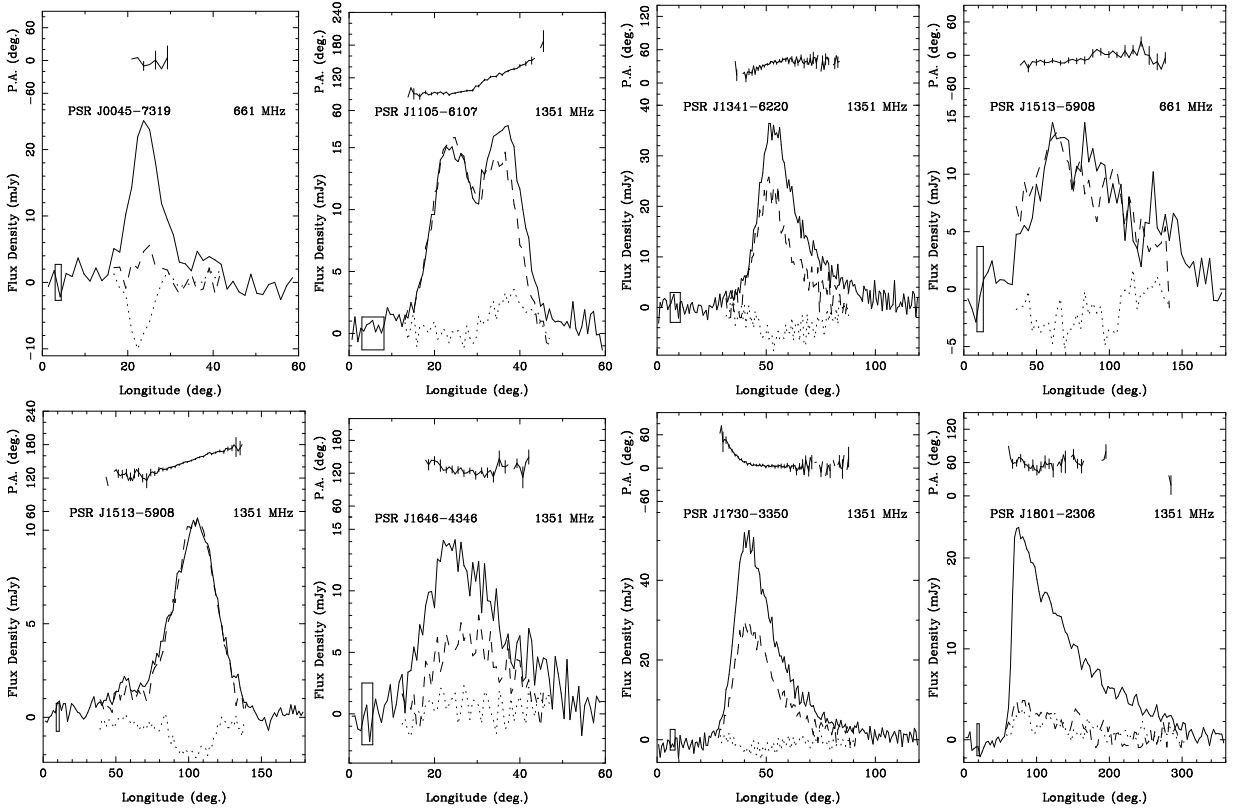} 
\caption{Polarization profiles for pulsars showing significant
polarization. In the lower half of each plot, the solid line indicates
total intensity as a function of pulse phase in degrees. The total
intensity has been normalized to $S$, the mean flux density, using
observations of Hydra A. The dashed and dotted lines indicate linearly
and circularly polarized intensity respectively. Positive values of
circular polarization are left-circularly polarized.  The height of
the box in the lower left-hand corner is twice the root-mean-square
(rms) scatter, $\sigma$, in the profile baseline. The upper half of
each plot shows the position angle as a function of pulse phase. The
pulsar name and center frequency of the observation are also indicated
in each plot.}
\label{fig-1} 
\end{figure}

\begin{figure}
\plotone{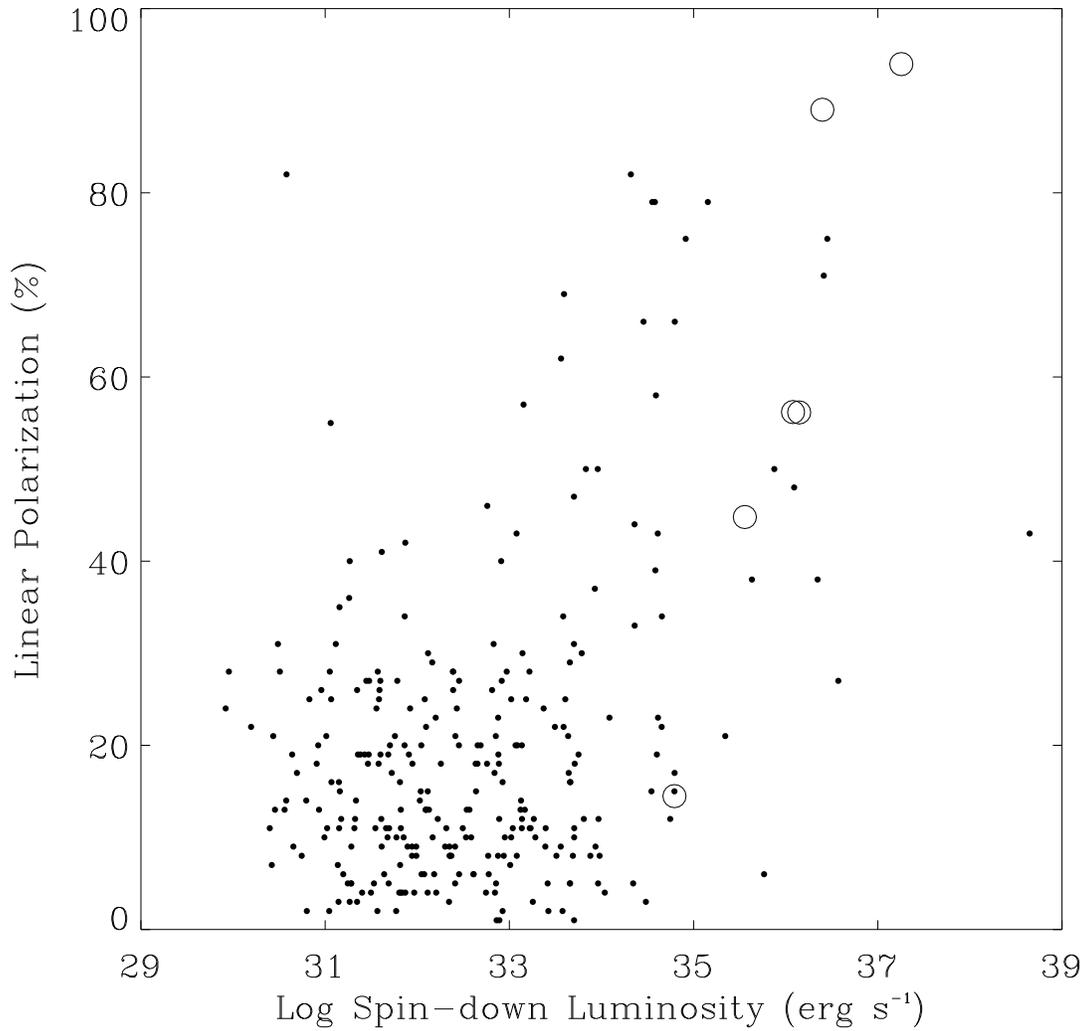} 
\caption{Linear polarization as a function of log $\dot{E}$ for the six
young pulsars in our sample at 1350 MHz (open circles) and 278 pulsars from
the 1400 MHz data of Gould \& Lyne (1998) (dots). There is a clear trend toward
stronger linear polarization at high $\dot{E}$ in both samples.}
\label{fig-2} 
\end{figure}


\begin{thebibliography}{}
 
\bibitem[Brazier \& Becker 1997]{bb97}
Brazier, K. T. S. \& Becker, W. 1997, \mnras, 284, 335

\bibitem[Camilo et al. 2000]{ckl+00} Camilo, F., Kaspi, V. M., Lyne,
A. G., Manchester, R. N., Bell, J. F., D'Amico, N., McKay, N. P. F.,
\& Crawford, F. 2000, \apj, 541, 367
     
\bibitem[Condon, Griffith, \& Wright 1993]{cgw93}
Condon, J. J., Griffith, M. R., \& Wright, A. E. 1993, \aj, 106, 1095

\bibitem[D'Amico et al. 1998]{dsb+98} 
D'Amico, N., Stappers, B. W., Bailes, M., Martin, C. E.,
Bell, J. F., Lyne, A. G., \& Manchester, R. N. 1998, \mnras,
297, 28 
       
\bibitem[Frail, Kulkarni, \& Vasisht 1993]{fkv93} 
Frail, D. A., Kulkarni, S. R., \& Vasisht, G. 1993,
\nat, 365, 136  
 
\bibitem[Gaensler et al. 1999]{gbm+99} 
Gaensler, B. M., Brazier, K. T. S., Manchester, 
R. N., Johnston, S., \& Green, A. J. 1999, \mnras, 305, 724. 

\bibitem[Gil, Kijak, \& Seiradakis 1993]{gks93}
Gil, J. A., Kijak, J., \& Seiradakis, J. H. 1993, \aap, 272, 268

\bibitem[Gould \& Lyne 1998]{gl98} 
Gould, D. M. \& Lyne, A. G. 1998, \mnras, 301, 235 

\bibitem[Gray et al. 1999]{gld+99}
Gray, A. D., Landecker, T. L., Dewdney, P. E., Taylor, A. R., Willis, A. G.,
\& Normandeau, M. 1999, \apj, 514, 221

\bibitem[Johnston et al. 1995]{jml+95} 
Johnston, S., Manchester, R. N., Lyne, A. G., Kaspi, V. M., \&
D'Amico, N. 1995, \aap, 293, 795

\bibitem[Kaspi et al. 1992]{kmj+92} 
Kaspi, V. M., Manchester, R. N., Johnston, S., Lyne, A. G.,
\& D'Amico, N. 1992, \apj, 399, L155 

\bibitem[Kaspi et al. 1993]{klm+93} 
Kaspi, V. M., Lyne, A. G., Manchester, R. N., Johnston, S.,
D'Amico, N., \& Shemar, S. L. 1993, \apj, 409, L57 

\bibitem[Kaspi et al. 1994a]{kms+94} 
Kaspi, V. M., Manchester, R. N., Siegman, B., Johnston, S., \&
Lyne, A. G. 1994a, \apjl, 409, L57 

\bibitem[Kaspi et al. 1994b]{kjb+94} 
Kaspi, V. M., Johnston, S., Bell, J. F., Manchester, R. N., Bailes,
M., Bessell, M., Lyne, A. G., D'Amico, N. 1994b, \apj, 423, L43

\bibitem[Kaspi et al. 1996a]{ktm96} 
Kaspi, V. M., Tauris, T. M., \& Manchester, R. N. 1996a, \apj, 459, 717

\bibitem[Kaspi et al. 1996b]{kmj+96} 
Kaspi, V. M., Manchester, R. N., Johnston, S., 
Lyne, A. G., \& D'Amico, N. 1996b, \aj, 111, 2028 

\bibitem[Kaspi et al. 1997]{kbm+97} 
Kaspi, V. M., Bailes, M., Manchester, R. N., Stappers, B. W., Sandhu,
J. S., Navarro, J., \& D'Amico, N. 1997, \apj, 485, 820 

\bibitem[Lyne \& Manchester 1988]{lm88} 
Lyne, A. G. \& Manchester, R. N. 1988, \mnras, 234, 477

\bibitem[Lyne, Pritchard, \& Graham-Smith 1993]{lpg93}
Lyne, A. G., Pritchard, R. S., \& Graham-Smith, F. 1993,
\mnras, 265, 1003 

\bibitem[Manchester \& Taylor 1977]{mt77} 
Manchester, R. N. \& Taylor J. H. 1977, Pulsars, San Francisco: Freeman 

\bibitem[Manchester et al. 1973]{mtt+73}
Manchester, R. N., Tademaru, E., Taylor, J. H., \& Huguenin,
G. R. 1973, \apj, 185, 951

\bibitem[Manchester, Tuohy, \& D'Amico 1982]{mtd82} 
Manchester, R. N., Tuohy, I. R., \& D'Amico, N. 1982, \apj, 262, L31

\bibitem[Manchester, D'Amico, \& Tuohy 1985]{mdt85} 
Manchester, R. N., D'Amico, N., \& Tuohy, I. R. 1985,
\mnras, 212, 975 

\bibitem[Manchester et al. 1991]{mkj+91} 
Manchester, R. N., Kaspi, V. M., Johnston, S., 
Lyne, A. G., \& D'Amico, N. 1991, \mnras, 253, 7P

\bibitem[Manchester 1996]{man97} 
Manchester, R. N. 1996, in IAU Colloq. 160, Pulsars: Problems and
Progress, ed. S. Johnston, M. Bailes, \& M. Walker (ASP
Conf. Ser. 105; San Francisco: ASP), 193

\bibitem[Manchester, Han, \& Qiao 1998]{mhq98} 
Manchester, R. N., Han, J. L., \& Qiao, G. J. 1998, \mnras, 295, 280

\bibitem[McConnell et al. 1991]{mmh+91} 
McConnell, D., McCulloch, P. M., Hamilton, P. A., Ables, J. G., Hall,
P. J., Jacka, C. E., \& Hunt, A. J. 1991, \mnras, 249, 654

\bibitem[Melatos 1997]{m97} 
Melatos, A. 1997, \mnras, 288, 1049

\bibitem[Navarro 1994]{n94} 
Navarro, J. 1994, PhD thesis, California Institute of Technology

\bibitem[Navarro et al. 1997]{nms+97}
Navarro, J., Manchester, R. N., Sandhu, J. S., Kulkarni, S. R.,
\& Bailes, M. 1997, \apj, 486, 1019

\bibitem[Qiao et al. 1995]{qml+95} 
Qiao, G. J., Manchester, R. N., Lyne, A. G., \& Gould D. M. 1995,
\mnras, 274, 572

\bibitem[Radhakrishnan \& Cooke 1969]{rc69} 
Radhakrishnan, V. \& Cooke, D. J. 1969, \aplett, 3, 225  

\bibitem[Rankin 1990]{r90} 
Rankin, J. M. 1990, \apj, 352, 247 

\bibitem[Rankin 1993a]{r93a} 
Rankin, J. M. 1993a, \apj, 405, 285 

\bibitem[Rankin 1993b]{r93b}
Rankin, J. M. 1993b, \apjs, 85, 145

\bibitem[Rankin \& Rathnasree 1997]{rr97}
Rankin, J. M. \& Rathnasree, N. 1997, 
Journal of Astrophysics and Astronomy, 18, 91

\bibitem[Romani \& Yadigaroglu 1995]{ry95} 
Romani, R. W. \& Yadigaroglu, I.-A. 1995, \apj, 438, 314

\bibitem[Seward \& Harnden 1982]{sh82} 
Seward, F. D. \& Harnden, F. R., Jr. 1982, \apj, 256, L45

\bibitem[Taylor \& Cordes 1993]{tc93} 
Taylor, J. H. \& Cordes, J. M. 1993, \apj, 411, 674

\bibitem[von Hoensbroech et al. 1998]{hkk98} 
von Hoensbroech, A., Kijak, J., \& Krawczyk, A. 1998, \aap, 
334, 571

\bibitem[Weisberg et al. 1999]{wcl+99}
Weisberg, J. M. et al. 1999, \apjs, 121, 171

\bibitem[Wu et al. 1993]{wml+93}
Wu, X., Manchester, R. N., Lyne, A. G., \& Qiao, G. 1993, \mnras, 261,
630

\end{thebibliography}
\end{document}